\documentclass[aps,pre,reprint,superscriptaddress, amsmath, amssymb, showpacs]{revtex4-2}

\usepackage{appendix}

 \usepackage{gensymb}
 \usepackage{subcaption}
 \usepackage{caption}
\newcommand*{\rom}[1]{\expandafter\@slowromancap\romannumeral #1@}

\usepackage{graphicx}
\usepackage{color}
\usepackage{xcolor}
\usepackage{comment} 

\begin{document}

\title{Molecular Dynamics Simulations of Temperature Relaxation in Non-Neutral Plasmas Relevant to Antimatter Experiments}



\author{James C. Welch III}
\affiliation{Nuclear Engineering \& Radiological Sciences, University of Michigan, Ann Arbor, Michigan 48109, USA}

\author{Louis Jose}
\affiliation{Nuclear Engineering \& Radiological Sciences, University of Michigan, Ann Arbor, Michigan 48109, USA}

\author{Timothy D. Tharp}
\affiliation{Physics, Marquette University, Milwaukee, Wisconsin, 53233, USA}

\author{Scott D. Baalrud}
\email{baalrud@umich.edu} 
\affiliation{Nuclear Engineering \& Radiological Sciences, University of Michigan, Ann Arbor, Michigan 48109, USA}


\date{\today}

\begin{abstract}
    An important process for antimatter experiments is the cooling of particles in a Penning-Malmberg trap to experimentally useful temperatures. A non-neutral plasma of one species (e.g. antiprotons) can be collisionally cooled on another colder species (e.g. electrons).
    Modeling temperature relaxation in these devices is challenging from a plasma physics perspective because the particles are strongly magnetized (the gyrofrequency exceeds the plasma frequency). Recently, a theoretical model was proposed to describe the temperature evolution in these conditions, predicting a multistep relaxation process where temperatures parallel to the magnetic field relax much faster than perpendicular to it. Here, this model is tested using molecular dynamics simulations. Two analysis methods are applied: one based on an imposed temperature difference, and the other based on a Green-Kubo relation. The results of the simulations support the theoretical predictions. This work extends previous studies of temperature anisotropy relaxation in one-component non-neutral plasmas to the two-component systems relevant to trapped antimatter experiments.
\end{abstract}




\maketitle


\section{Introduction}

Antimatter experiments such as the Antihydrogen Laser Physics Apparatus (ALPHA), manipulate non-neutral plasmas confined in Penning-Malmberg traps~\cite{Akbari2025_ALPHA2,Andresen_PRL_2008, Baker_Nature_Comm_2021, Fajans_POP_2020, Andresen2011}. 
This often involves mixing plasmas of different species to promote the collisional cooling of one species on another, for example, hot antiprotons cooling on cold electrons.
The density of the plasma in these traps must be below the Brillouin density limit in order to be confined, which makes them strongly magnetized in the sense that the gyrofrequency, $\omega_{c} = eB/m$, exceeds the plasma frequency, $\omega_{p} = \sqrt{n e^2 / m \epsilon_0}$, i.e., $\beta \equiv \omega_{c} / \omega_{p} > 1$~\cite{Davidson2001Book, Baalrud_PRE_2017}.
This is true for electrons $(\beta_e>1)$ and ions $(\beta_i>1)$.
Standard temperature relaxation models are not applicable under these conditions because a fundamental assumption of standard plasma kinetic theory is that $\beta_e <1$. 
A recent kinetic theory~\cite{josePRE2025,Jose_POP_2020,Jose_POP_2021,Jose_POP_2022,Jose_POP_2023} proposed that the plasma in these conditions undergo a two-step relaxation process where strong magnetization speeds up the relaxation rate parallel to the magnetic field but significantly slows it in the perpendicular direction. 
Here, we use molecular dynamics (MD) simulations to test this prediction and find rates consistent with the model. In addition to its application to the ALPHA experiment, this work also has implications for other strongly magnetized systems such as dusty plasmas~\cite{Thomas_PPCF_2012}, non-neutral plasmas~\cite{Beck_PRL_1992,beck1996temperature,Glinsky_Phys_Fluids_1992}, magnetized ultracold neutral plasmas~\cite{Gorman_PRL_2021, Gorman_PRA_2022, Zhang_PRL_2008, Sprenkle_PRE_2022, Guthrie_POP_2021, Pak_PRE_2024} and pinch experiments~\cite{Bennett_Phys_Rev_Accel_Beams_2021}.

Traditional plasma kinetic theories assume that the plasma is weakly magnetized ($\beta_e,\beta_i\ll1)$ and, as a consequence, predict that temperature relaxation rates are independent of the magnetic field strength~\cite{ichimaru2018statistical}. 
In this situation, electron-electron and ion-ion collision rates are predicted to be much faster than the ion-electron collision rate due to their similar masses. 
This implies that if electrons and ions start with different temperatures, the velocity distribution of each species tends to remain isotropic and characterized by a species-specific temperature as the two temperatures evolve to an equilibrium value. 
Similarly, if a temperature anisotropy is imposed, as occurs in wave heating~\cite{Stix1992}, the anisotropy relaxation rate is expected to be much faster than the interspecies temperature relaxation rate. 

The situation in strongly magnetized plasmas is different. 
Previous experimental and theoretical works in single species plasmas~\cite{Beck_PRL_1992,beck1996temperature,Glinsky_Phys_Fluids_1992,jose_submitted_2025,hyatt1987measurement,Baalrud_PRE_2017,Hjorth_Phys_Fluids_1987,Oneil1990,Oneil_Phys_Fluids_1983} have shown that the relaxation of a temperature anisotropy is severely suppressed by strong magnetization. 
Recently, a model~\cite{josePRE2025} was developed to describe the temperature evolution for a two-component (electron-ion) plasmas in strongly magnetized conditions. 
This predicts that strong magnetization generally increases collision rates parallel to the magnetic field, while decreasing them perpendicular to the magnetic field. 
This asymmetry in the collision rates
fails to suppress and can temporarily generate
anisotropies during the electron-ion equilibration process. 
Furthermore, the perpendicular relaxation rates can be so severely suppressed that the anisotropy relaxation time, particularly for electrons, sets the overall temperature relaxation rate (it is the slowest process). 
Since it takes two temperatures (parallel and perpendicular) to describe the evolution of each species, it was found that six independent relaxation rates must be specified in order to describe the full relaxation process. 

Here, we test this model using first-principles MD simulations. Two complementary simulation methods are used. 
The first is a non-equilibrium setup (MD-N), where an initial temperature difference between ions and electrons is created and the system is allowed to relax freely. Relaxation rates are obtained by fitting an exponential to the time-dependent temperatures. Because this method relies on observing and fitting relaxation rates, only the dominant rates can be extracted at a fixed set of conditions.  This method is found to be effective in extracting two of the six relaxation rates. 
The second method (MD-GK) conducts a simulation at thermodynamic equilibrium and analyzes fluctuations in the temperatures using Green-Kubo relations. This method offers an advantage in that it allows direct access to each of the relaxation rates, in principle. 
In practice, two of the six coefficients were too small compared to the simulation noise to be resolved.
It is also significantly more computationally expensive than the MD-N method. 

The results of the MD simulations generally agree well with the model predictions. 
The temperature evolution [Fig.~\ref{fig: MD_theo}] shows good overall agreement, including a shortened parallel relaxation and prolonged perpendicular relaxation for electrons at strong magnetization. 
Similarly, ion-electron relaxation rates [Fig.~\ref{fig: nuie00}] extracted from both simulation methods (MD-N and MD-GK) indicate an increased collision rate in the parallel direction at strong magnetization.
Furthermore, electron isotropization rates from MD-GK [Fig.~\ref{fig: nuee_NEMD_TCP}] show suppression by orders of magnitude in the strongly magnetized regime. Along with this suppression, electron isotropization rates also exhibit a clear contribution from electron-ion collisions in addition to electron-electron collisions~\cite{Baalrud_CPP_2017, Baalrud_PRE_2017, jose_submitted_2025}. 
Ion isotropization rates [Fig.~\ref{fig: nuii_TCP}] below $\beta_i<1$ are consistent with the unmagnetized expectation.
The MD-GK relaxation rates were determined by correlating the fluctuations of temperature differences [Fig.~\ref{fig: corrs}].
Also, consistent with model predictions, the presence of an ion-electron temperature difference generates an electron anisotropy as observed in the MD-GK simulations [Fig.~\ref{fig: rho_ie_01}]. At high magnetization, the overall relaxation process is characterized by a relatively fast relaxation between the ion and parallel electron temperatures, while the perpendicular electron temperature takes orders of magnitude longer come to equilibrium with the rest.

Previous works have also studied temperature relaxation in plasmas using MD simulations. 
Dimonte and Daligault used the non-equilibrium method (MD-N) to investigate ion-electron equilibration in unmagnetized plasmas~\cite{Dimonte_PRL_2008}. 
Isotropization in weakly magnetized~\cite{Baalrud_CPP_2017} and strongly magnetized~\cite{Baalrud_PRE_2017} one-component plasmas (OCP) was studied using the MD-N method, and recent work applied the MD-GK method to this problem as well~\cite{jose_submitted_2025}.  
Other works in unmagnetized plasmas have obtained temperature relaxation rates from the stopping power of a single ion moving through an electron plasma~\cite{bernstein2022method,David_POP_2019}. These previous works considered only one-component plasmas in the strongly magnetized regime, or only the unmagnetized regime for ion-electron plasmas.
Furthermore, each of these has focused on one relaxation process at a time. In contrast, this work addresses the complexity of the temperature relaxation process by calculating multiple relaxation rates that simultaneously occur in strongly magnetized ion-electron plasmas. 

Testing this theory is relevant to the ALPHA experiment~\cite{Andresen_PRL_2008, Baker_Nature_Comm_2021, Fajans_POP_2020,baker2025precision, Ahmadi_nature_comm_2017}. Cold plasma temperatures are necessary for the production of trappable antihydrogen atoms, and this is often achieved through sympathetic cooling between plasma species~\cite{GlinskyONeil1991,butler2011thesis}. For example, laser-cooled Beryllium ions are used to sympathetically cool positrons to very low temperatures for more efficient antihydrogen production~\cite{Baker_Nature_Comm_2021}. The theory of temperature relaxation tested here~\cite{josePRE2025} can model this application as well as antiprotons cooling on electrons.

\section{Rate Equations \label{sec: rate_eqns}}

In a weakly magnetized plasma, energy relaxation by intraspecies collisions (e-e or i-i) is faster than by interspecies collisions (e-i) by at least a factor of mass ratio $m_i/m_e$ for electrons and the square root of the mass ratio $\sqrt{m_i/m_e}$ for ions. 
For this reason, any imposed temperature anisotropy is expected to relax at a much faster rate than the electron-ion temperature equilibration. 
The ion and electron distribution functions can then be modeled as isotropic, and the associated temperatures evolve according to
\begin{subequations}
\label{eq: unmag_temp_evo}
\begin{eqnarray}
\label{eq:Ti1}
    \frac{d T_{i}}{d t} &=& - \nu_o^{ie} (T_i -T_e) \\
    \label{eq:Te1}
    \frac{d T_{e}}{d t} &=& - \nu_o^{ei} (T_e -T_i).
\end{eqnarray}
\end{subequations}
Here, $T_i$ is the ion temperature, $T_e$ is the electron temperature, and 
\begin{equation}
\label{eq:classical_nu}
    \nu_o^{ss^\prime} = \frac{32 \sqrt{\pi} q_s^2 q_{s'}^2 n_{s'} \ln \left( \Lambda_{ss'} \right)}{3 m_s m_{s'} \left( v_{Ts}^2 + v_{Ts'}^2 \right)^{3/2}}
\end{equation}
is the temperature relaxation rate between species $s$ and $s^\prime$, where $q_s$ is the charge, $m_s$ is the mass, $n_s$ is the number density, $v_{Ts} = \sqrt{2 k_B T_s / m_s}$ is the thermal speed, and $\ln \left( \Lambda_{ss'} \right)$ is the Coulomb logarithm. When plasma enters the strong Coulomb coupling regime, the Coulomb logarithm diverges. Previous works have generalized the Coulomb logarithm to strong coupling~\cite{Baalrud_PRL_2013,Baalrud_POP_2014}.

Energy conservation $d(T_e+T_i)/dt=0$ implies that the rates must be equal: $\nu_o^{ie}=\nu_{o}^{ei}$. 
Subtracting Eqs.~(\ref{eq:Ti1}) and (\ref{eq:Te1}) then provides a single equation for the temperature difference: \
\[\frac{d\Delta T}{dt} = -2 \nu^{ie}_o \Delta T,\] 
where $\Delta T=T_i-T_e$.

The situation is qualitatively different when a plasma is strongly magnetized, because in this regime collision rates are dependent on direction (with respect to the magnetic field). 
This leads to a situation where temperature anisotropies can form during the relaxation process. 
Furthermore, the intraspecies anisotropy relaxation rate, particularly for electrons, is exponentially suppressed ~\cite{Beck_PRL_1992,beck1996temperature,Glinsky_Phys_Fluids_1992,jose_submitted_2025,hyatt1987measurement,Baalrud_PRE_2017,Hjorth_Phys_Fluids_1987}. 
So even though the weakly magnetized e-e collision rate is faster than the e-i rate by a factor of $m_i/m_e$ due to the mass scaling in Eq.~(\ref{eq:classical_nu}), the magnetic field can suppress this so much that a long-lived electron temperature anisotropy can persist. 
Similar arguments apply to ions when the ion magnetization parameter is large $\beta_i\gg 1$.  

Since each species has distinct temperatures parallel and perpendicular to the magnetic field, the system is defined by four temperatures: the parallel ion temperature $(T_{i\parallel})$, the perpendicular ion temperature $(T_{i\perp})$, the parallel electron temperature $(T_{e\parallel})$ and the perpendicular electron temperature $(T_{e\perp})$. 
These can be defined as moments of the distribution function 
\begin{subequations}
\begin{align}
T_{s\parallel} &= \frac{1}{n_s} \int d^3v\,  m _s v_\parallel^2 f_s(\mathbf{v}) , \\
T_{s\perp} &= \frac{1}{n_s} \int d^3v\,  \frac{1}{2} m _s v_\perp^2 f_s(\mathbf{v}) ,    
\end{align}
\end{subequations}
such that $T_s = \frac{1}{3} T_{s\parallel} + \frac{2}{3} T_{s\perp}$. 
Evolving an electron-ion system then depends on four rate equations and 12 rate coefficients $\nu^{\alpha \beta}$ of the form
\begin{equation}
\label{eq:gen_rate}
    \frac{dT_\alpha}{dt} = - \sum_{\beta\neq \alpha} \nu^{\alpha \beta} (T_\alpha - T_\beta),
\end{equation}
where $\alpha$ and $\beta$ index the four temperatures $T_{i\parallel}$, $T_{i\perp}$, $T_{e\parallel}$ and $T_{e\perp}$. 

Energy conservation implies relations between three pairs of relaxation rates~\cite{josePRE2025}. Similar to the weakly magnetized case, the number of evolution equations is reduced from four to three by formulating the evolution equations in terms of temperature differences: the temperature difference between ions and electrons $(T_i - T_e)$, the ion anisotropy $(\Delta T_i \equiv T_{i\parallel} - T_{i\perp})$, and the electron anisotropy $(\Delta T_e \equiv T_{e\parallel} - T_{e\perp})$. This formulation is characterized by only $9$ rates, with the temperature evolution given as
\begin{subequations}\label{eq:T_evolution}
\begin{eqnarray}
\label{eq:T_ie_evol}
    \frac{d \Delta T_{ie}}{d t} &=& - \eta^{ie} \Delta T_{ie}+ \rho^{ie}_{i} \Delta T_i + \rho^{ie}_{e} \Delta T_e , \\
    \label{eq:T_i_evol}
    \frac{d \Delta T_i}{d t} &=& \rho^{i}_{ie} \Delta T_{ie} - \eta^{i} \Delta T_i + \rho^{i}_{e} \Delta T_e , \\
    \label{eq:T_e_evol}
    \frac{d \Delta T_e}{d t} &=&  \rho^{e}_{ie} \Delta T_{ie}
    + \rho^{e}_{i} \Delta T_i 
    - \eta^{e} \Delta T_e  .
\end{eqnarray} 
\end{subequations}
Here, the ion-electron temperature difference is defined as $\Delta T_{ie} \equiv \frac{3}{2} (T_i - T_e)$. The factor of 3/2 is included to construct a symmetric form of the rate matrix $\mathcal{R}$. This symmetry can be shown from general arguments of the evolution equations in a linear regime, as described in Appendix~\ref{app: C_sym}.
It implies that the superscripts and subscripts of $\rho$ are interchangeable, e.g., $\rho^{ie}_{i} = \rho^{i}_{ie}$, $\rho^{ie}_{e} = \rho^{e}_{ie}$, and $\rho^{i}_{e} = \rho^{e}_{i}$. Thus, the evolution is fully characterized by only $6$ independent rates. 
Alternatively, Eq.~(\ref{eq:T_evolution}) can be written in matrix form as
\begin{eqnarray}
    \label{eq: NM_temp_evo}
    \frac{d \Delta \textbf{T}}{dt} = -\mathcal{R} \cdot \Delta \textbf{T} ,
\end{eqnarray}
where the temperature difference vector is defined as
\begin{eqnarray}
    \Delta \textbf{T} \equiv \begin{bmatrix}
        \Delta T_{ie}\\
        \Delta T_i\\
        \Delta T_e 
    \end{bmatrix} ,
\end{eqnarray}
and the symmetric rate matrix is defined as
\begin{eqnarray}
\label{eq: nu_matrix}
    \mathcal{R} \equiv 
    \begin{bmatrix}
        \eta^{ie} & - \rho^{ie}_{i} & - \rho^{ie}_{e}\\
        - \rho^{i}_{ie} & \eta^{i} & - \rho^{i}_{e} \\
        - \rho^{e}_{ie} & - \rho^{e}_{i} & \eta^{e} 
    \end{bmatrix} .
\end{eqnarray}

In Eqs.~(\ref{eq:T_evolution}) and (\ref{eq: nu_matrix}), relaxation rates $(\eta)$ are differentiated from coupling rates $(\rho)$. The distinction is that relaxation rates represent the intrinsic decay of a temperature difference, and therefore have a positive definite sign, whereas coupling rates quantify how one temperature difference influences another, and may have either a positive or negative sign. The physical meaning of the sign of the coupling rates is discussed in Sec.~\ref{sec: EMD_rates}.
In the notation for $\eta$, the superscript denotes the temperature difference being relaxed, where ``$ie$'' is the ion-electron temperature difference, ``$i$'' is the ion anisotropy, and ``$e$'' is the electron anisotropy. 
Kinetic theory, unlike the 12-rate and 9-rate descriptions, reveals that anisotropy relaxation rates consist of contributions from intraspecies and interspecies collisions. Specifically,
\begin{subequations}\label{eq: anisotropy_decomp_terms}
\begin{eqnarray}
    \eta^{i} &=& \eta^{i,i|i} + \eta^{i,i|e} , \label{eq: i_anisotropy_decomp_terms} \\
    \eta^{e} &=& \eta^{e,e|e} + \eta^{e,i|e} ,
    \label{eq: e_anisotropy_decomp_terms}
\end{eqnarray} 
\end{subequations}
where $\eta^{i,i|i}$ and $\eta^{e,e|e}$ arise from ion–ion and electron–electron collisions, respectively, and $\eta^{i,i|e}$ and $\eta^{e,i|e}$ arise from ion–electron collisions.
For example, $\eta^{i,i|e}$ physically represents the ion isotropization due to ion-electron collisions.
In terms of the coupling rates, $\rho^{l}_k$, the subscript ($k$) specifies the driving temperature difference and the superscript ($\l$) denotes the affected one. 
Since the notation here is different from that used in the model of Ref.~\cite{josePRE2025}, the relationships between the definitions are provided in Appendix \ref{app: def_rates}.

\section{Non-equilibrium MD \label{sec: NEMD}}

\subsection{Simulation Setup}

\begin{figure*} [!htb]
\centerline{\includegraphics[width = 1.80\columnwidth]{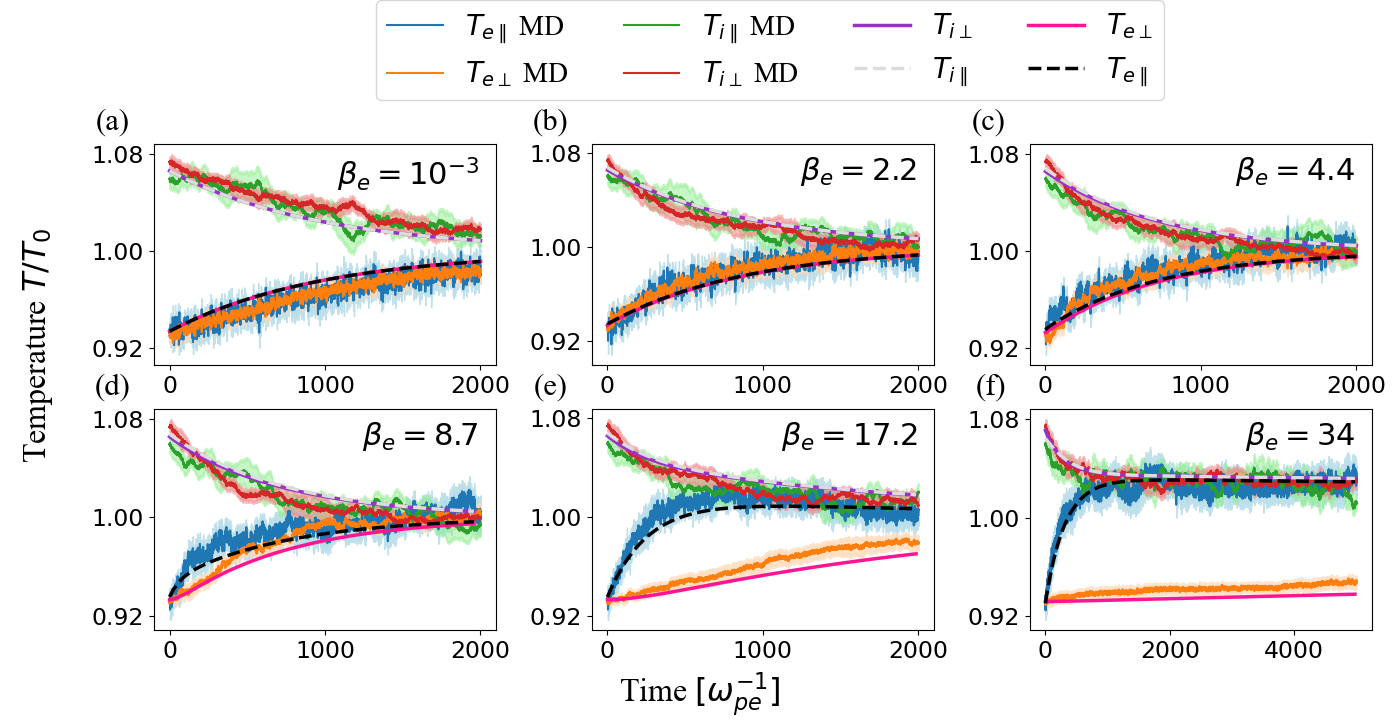}}
\caption {Temperature evolution from MD simulations (blue, orange, green, and red) averaged over several runs, compared with the numerical evaluation of the model from Ref.~\cite{josePRE2025} (violet, white, pink, and black). Shaded regions indicate the standard deviation from the run-to-run average.}
  \label{fig: MD_theo}
\end{figure*}

Molecular dynamics simulations were performed using the open-source software LAMMPS~\cite{LAMMPS}. In these simulations, all particles interact classically through a repulsive electrostatic Coulomb force. To be numerically efficient, the particle-particle particle-mesh method was used~\cite{frenkel2002molecular}. 
To simulate an infinite plasma, periodic boundary conditions were implemented in the cubic simulation domain. In each simulation, 5000 electrons and 5000 ions were used. 
All particles were assigned a positive charge. 
The mass ratio $(m_r \equiv m_i/m_e)$ was chosen to be $m_r=1000$ to correspond to previous MD simulations~\cite{David_PRE_2020, David_POP_2019, David_POP_2021}. Larger mass ratio simulations are more expensive since the ion-electron relaxation rate scales inversely with the mass ratio. 
In all simulations, the Coulomb coupling strength based on the electron density 
\begin{equation}
    \Gamma_e = \frac{e^2/a_e}{4\pi \epsilon_o k_BT}, 
\end{equation}
was set to $\Gamma_e=1$. 
Here, $a_e = (3/4\pi n_e)^{1/3}$ is the average inter-electron spacing. 
The magnetic field strength ($\beta_e$) was varied. 

To initialize the simulations, a pure electron plasma with no magnetic field was equilibrated using a Nos\'e-Hoover thermostat~\cite{frenkel2002molecular}. Then, the masses of half the electrons were increased by a factor of $1000$ to represent ions, and their velocities were scaled accordingly to maintain their temperature. Next, the velocities of the ions were increased and electrons decreased to introduce a $10\%$ temperature difference. 
Fast anomalous relaxation observed in previous studies~\cite{Baalrud_CPP_2017, Baalrud_PRE_2017} is attributed to this step, where the plasma is brought out of equilibrium.
The magnetic field was then applied. From the Bohr-van Leeuwen theorem~\cite{Pathria2011Book}, the plasma configuration was assumed to remain unchanged, as the system was near equilibrium.
The system was then transitioned to the microcanonical (NVE) ensemble and allowed to relax back to equilibrium. The parallel and perpendicular temperatures of each species were recorded during the NVE phase. 

To ensure that each simulation remained true to a first-principles approach, it was mandated that the smallest relevant motion be well resolved. In weakly magnetized simulations, this is the electron plasma period, while in strongly magnetized simulations it is the electron gyroperiod; thus, the timestep was chosen to be $t_s \leq \text{min} \{ \omega_{pe}^{-1}, \omega_{ce}^{-1} \} /100$. The duration of each simulation ranged from thousands to tens of thousands of plasma periods, depending on the time required to relax to equilibrium at a given magnetic field strength.

\subsection{Temperature Evolution \label{sec: temp_evo_results}}

The temperature evolution for a sample set of conditions is shown in Fig.~\ref{fig: MD_theo}. 
This illustrates the transition from an unmagnetized system that relaxes isotropically to a strongly magnetized one where the electron perpendicular temperature relaxation is heavily suppressed. 
It also shows a comparison with the model from Ref.~\cite{josePRE2025}, showing broad agreement. 
Figure~\ref{fig: MD_theo}(a) ($\beta_e = 0.001$) displays an effectively unmagnetized case, where it is observed that $T_{i\parallel}=T_{i\perp}$ and $T_{e\parallel} = T_{e\perp}$ through the evolution. 
As the magnetization parameter increases to $\beta_e = 2.2$ and $4.4$, (Figs.\ref{fig: MD_theo}(b-c)), the relaxation becomes more rapid. 
This is expected because strong magnetization causes the energy exchange rate between ions and electrons in the parallel direction to increase, while the even more rapid interspecies collisions maintain isotropic temperatures.

As the magnetization strength is further increased to $\beta_e = 8.7$ (Fig.\ref{fig: MD_theo}(d)), a weak electron anisotropy appears, indicating that the electron-electron anisotropy relaxation rate has become suppressed enough (in comparison to the electron-ion relaxation rate) to overcome the mass ratio factor in Eq.~(\ref{eq:classical_nu}). 
This prolongs the overall relaxation. 
In Fig.~\ref{fig: MD_theo}(e) ($\beta_e = 17.2$), the suppression of the perpendicular electron temperature relaxation is clearly visible: the parallel electron temperature relaxes much faster than the perpendicular component. By $\beta_e = 34$ (Fig.~\ref{fig: MD_theo}(f)), the electron isotropization is so heavily suppressed that the overall relaxation is extended by orders of magnitude. The virtually unchanging perpendicular electron temperature in Fig.~\ref{fig: MD_theo}(f) provides clear evidence that it is an adiabatic invariant. 
Similar observations have previously been made in single component plasmas~\cite{Oneil_Phys_Fluids_1985}; and here is observed in a two component system.

\subsection{Reduced models \label{sec: red_mods}}

To measure relaxation rates, the time-dependent temperatures were fit to exponential functions. However, individual relaxation rates can only be identified this way if the  associated temperature evolution equation is described by a single rate, which is not true in the general evolution described by Eq.~(\ref{eq:T_evolution}). 
Nevertheless, the general equations can be decoupled in certain limits, and a subset of the relaxation rates identified in these limits. 

From the unmagnetized limit transitioning into the strongly magnetized regime, intraspecies energy exchange dominates over interspecies energy exchange. This prevents the development of any temperature anisotropy. Consequently, Eq.~(\ref{eq:T_ie_evol}) reduces to a form similar to  Eq.~(\ref{eq: unmag_temp_evo}), 
\begin{eqnarray}
\label{eq: unmag_rm}
\frac{d\Delta T_{ie}}{dt} = - \eta^{ie} \Delta T_{ie} ,
\end{eqnarray}
but with a relaxation rate $\eta^{ie}$ that depends on the magnetic field strength. 

As confirmed by the temperature evolution shown in Fig.~\ref{fig: MD_theo}, 
this behavior persists up to $\beta_e \approx 5$, beyond which an electron anisotropy becomes apparent. We refer to this as the isotropic reduced model, since both species remain isotropic throughout the evolution. Based on Eq.~(\ref{eq: unmag_rm}), the electron-ion relaxation rate $\eta^{ie}$ is determined by fitting $\Delta T_{ie}$ to an exponential when $\beta_e \lesssim 5$. 

At a large magnetization strength ($\beta_e \gtrsim 30$), electron intraspecies energy exchange is so strongly suppressed that the electron isotropization occurs on a timescale much longer than the ion-electron temperature relaxation~\cite{josePRE2025}. In contrast, ion isotropization remains fast when $\beta_e \lesssim 130$, keeping the ion distribution effectively isotropic. 
As seen in Fig.~\ref{fig: MD_theo}, isotropic ions equilibrate with the parallel electron temperature on a timescale faster than the electron isotropization. 
Assuming $T_{e\perp}$ is constant over this timescale, using energy conservation, and noting that in this regime $\eta^{ie}=\rho^{ie}_e$, Eq.~(\ref{eq:T_ie_evol}) reduces to
\begin{eqnarray} \label{eq: intermed_rm}
\frac{d(T_i - T_{e\parallel})}{dt} = - 2 \eta^{ie}(T_i-T_{e\parallel}) . 
\end{eqnarray}
In this regime ($30 \lesssim \beta_e \lesssim 130$), $\eta^{ie}$ is obtained by fitting the early time evolution of $T_i - T_{e\parallel}$ to an exponential decay.

At large $\beta_e$, the slowest process is the equilibration of the electron perpendicular temperature. Assuming $T_i = T_{e\parallel}$ and using energy conservation, Eq.~(\ref{eq:T_ie_evol}) and~(\ref{eq:T_e_evol})  reduce to
\begin{eqnarray}
    \frac{\Delta T_e}{dt} = - \frac{1}{2} \Bigl(
    \eta^e
    - \rho^{ie}_{e} \Bigr) \Delta T_e .
    \label{eq: DeltadTe_slow_full}
\end{eqnarray}
In this regime, the intraspecies rate dominates $( \eta^{e,i|e} - \rho^{ie}_{e} \ll \eta^{e,e|e})$, allowing the model to be approximated as
\begin{eqnarray}
\frac{\Delta T_e}{dt} = - \frac{1}{2} \eta^{e,e|e} \Delta T_e .
\label{eq: DeltadTe_slow}
\end{eqnarray}
The isotropization rate $\eta^{e,e|e}$ is extracted by fitting the late-time evolution of $\Delta T_e$ to an exponential. Because electron isotropization occurs slowly, simulations must run for extended times, limiting how many rates can be computed due to computational cost.

Relaxation rates cannot be determined from MD-N for magnetization strengths between $5 \lesssim \beta_e \lesssim 30$, as no reduced models are valid in this regime. Furthermore, a maximum of two relaxation rates can be determined when the reduced models are valid. The MD-GK method has no such limitations. It can, in principle, access all magnetization strengths and determine four of the six rates. However, they are more computationally expensive and this can be a limitation.

\subsection{Relaxation rates \label{sec: NEMD_results}}

\begin{figure} 
\centerline{\includegraphics[width = 8.5cm]{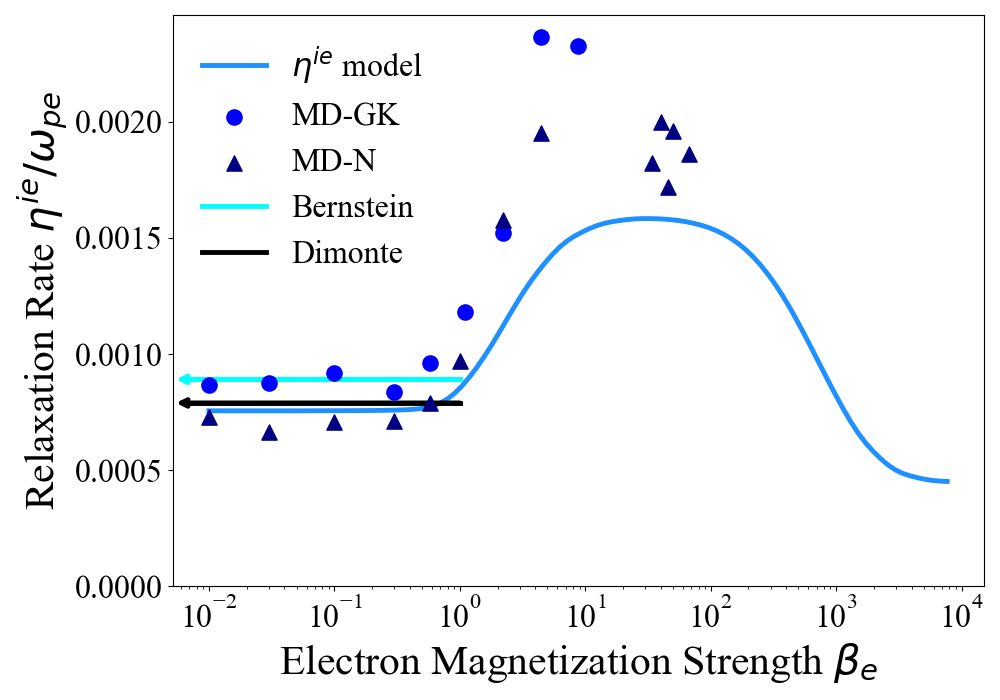}}
\caption {The ion-electron relaxation rate $(\eta^{ie})$ is displayed as a function of electron magnetization strength $(\beta_e)$. The blue curve represents the theory. The triangles are from MD-N. The circles are from MD-GK. The cyan arrow shows a previous unmagnetized stopping power MD-N simulation~\cite{bernstein2022method}. The black arrow shows a previous unmagnetized MD-N simulation~\cite{Dimonte_PRL_2008}.}
  \label{fig: nuie00}
\end{figure}

Figure~\ref{fig: nuie00} shows the electron-ion temperature relaxation rate ($\eta^{ie}$) as a function of the electron magnetization strength. 
The results are consistent with a constant value independent of magnetic field strength until the strong magnetization regime is reached ($\beta_e \gtrsim 1$), at which point the relaxation rate increases by approximately a factor of 2.5, before plateauing when $\beta_e \gtrsim 10$. 
The value obtained in the weak magnetization cases is consistent with previous works by Dimonte and Daligault~\cite{Dimonte_PRL_2008} and Bernstein \emph{et al}~\cite{bernstein2022method} that considered unmagnetized plasmas. 
A comparison with the model predictions from~\cite{josePRE2025} shows agreement with the general trend that the relaxation rate increases by approximately a factor of 2.5 before plateauing in the $10\lesssim \beta_e \lesssim 100$ range. 
There is clearly some quantitative difference of approximately $\sim 20$\% between the simulation results and the model calculations in the plateau region. 

\begin{figure} [!htb] 
\centerline{\includegraphics[width = 8.5cm]{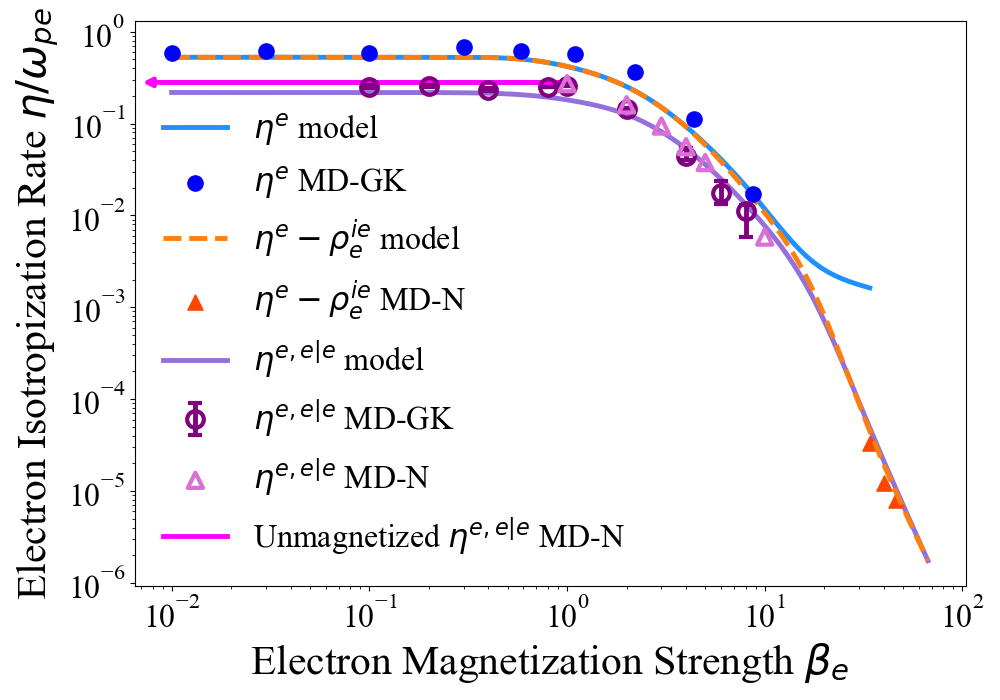}}
\caption {Three rates related to electron isotropization are shown: the electron isotropization rate $(\eta^e)$ (blue), the electron isotropization rate due to electron-electron collisions $(\eta^{e,e|e})$ (purple-pink), and the rate predicted by the reduced model in Eq.~(\ref{eq: DeltadTe_slow_full}) ($\eta^{e}- \rho^{ie}_{e}$) (orange). Model predictions are shown as solid or dashed lines. Filled in markers display new ion-electron MD data, where MD-GK are circles and MD-N are triangles. Also shown with hollow markers are results of previously published work for a one-component plasma that used a similar MD-N method: Ref.~\cite{Baalrud_PRE_2017} (triangles) for a magnetized plasma, and Ref.~\cite{Baalrud_CPP_2017} (pink arrow) for an unmagnetized plasma, and results using an MD-GK method~\cite{jose_submitted_2025} (circles).}  \label{fig: nuee_NEMD_TCP}
\end{figure}

Figure~\ref{fig: nuee_NEMD_TCP} shows results for the electron anisotropy relaxation rate as a function of $\beta_e$, showing good agreement with the model predictions. 
Recall that as described in Eq.~(\ref{eq: DeltadTe_slow_full}), the MD-N simulations can only access the rate $\eta^{e}- \rho^{ie}_{e}$. 
However, the model predicts that $\eta^{e,e|e} \gg |\eta^{e,i|e}- \rho^{ie}_{e}|$, so that the electron-electron collision component dominates and the rate equation simplifies to Eq.~(\ref{eq: DeltadTe_slow}). 
The good agreement between the MD results and the model calculations that assume this rate to be dominant confirms this expectation. 
Since the electron-electron rate is expected to be largest, results are also shown from previous work on the one-component plasma that also computed this rate.~\cite{Baalrud_CPP_2017,Baalrud_PRE_2017,josePRE2025} 
Good agreement is observed across the broad range of magnetization strength. 
The new MD-N data is able to extend the MD results to a much larger $\beta_e$ value than the previous simulations. 
The striking feature is that the anisotropy relaxation rate decreases exponentially in the strongly magnetized regime, dropping by more than 4 orders of magnitude as $\beta_e$ ranges from 1 to 40. 
This implies that a temperature anisotropy can be extremely long-lived in a strongly magnetized plasma. 

\section{Green-Kubo MD \label{sec: EMD}}
\subsection{Green-Kubo Theory \label{sec: GreenKubo}}
\subsubsection{General Form \label{sec: GK_derivation}}

Green-Kubo relations relate correlations of the time-dependent fluctuations of a system in equilibrium to the linear transport coefficients associated with perturbations of the system away from equilibrium~\cite{green1954markoff,kubo1957statistical}.
In the MD-GK simulations, the temperature fluctuations of an equilibrium plasma are recorded and related to relaxation rates via Green-Kubo relations.
Here, a general Green-Kubo relation for all six relaxation and coupling rates associated with Eq.~(\ref{eq:T_evolution}) is derived following the method of Zwanzig~\cite{Zwanzig_1964}. In Sec.~\ref{sec: simp_GK}, approximate forms for relevant rates are provided.

The temperature evolution in Eq.~(\ref{eq: NM_temp_evo}) assumes a Markoffian approximation. At strong coupling or strong magnetization, this assumption fails. Thus, we generalize the model to a non-Markoffian form
\begin{equation}
\label{eq: M_temp_evo}
\frac{d \Delta \textbf{T} (t)}{dt} = -\int^t_0 \mathcal{R} (t-t') \cdot \Delta \textbf{T} (t') ,
\end{equation}
where $\mathcal{R}(t)$ is the time-dependent rate matrix. The relevant rates are obtained from the long-time (low-frequency) limit of $\mathcal{R}(t)$. To isolate these coefficients, first apply a Laplace transform to Eq.~(\ref{eq: M_temp_evo}), providing
\begin{equation}
\label{eq: GK_derive_step}
    i \omega \Delta \Tilde{\textbf{T}} (\omega) - \Delta \textbf{T} (0) 
    = - \Tilde{\mathcal{R}} (\omega) \cdot \Delta \Tilde{\textbf{T}} (\omega) .
\end{equation}
Next, taking the outer product of Eq.~(\ref{eq: GK_derive_step}) with the initial temperature difference vector, $\Delta \textbf{T}(0)$, then the ensemble average provides 
\begin{equation}
\label{eq: LP_nu_unsolved}
    i \omega \Tilde{\mathcal{C}} (\omega) - \mathcal{C} (0) 
    = - \Tilde{\mathcal{R}} (\omega) \cdot  \Tilde{\mathcal{C}} (\omega) 
\end{equation}
where
\begin{eqnarray}
\label{eq: corr_matrix}
    \mathcal{C} \equiv 
    \begin{bmatrix}
        C^{ie}_{ie} (t)  & C^{ie}_{i} (t) & C^{ie}_{e} (t)\\
        C^{i}_{ie} (t)  & C^{i}_{i} (t) & C^{i}_{e} (t) \\
        C^{e}_{ie} (t)  & C^{e}_{i} (t) & C^{e}_{e} (t)
    \end{bmatrix} 
\end{eqnarray}
is the matrix of time correlation functions 
\begin{equation}
\label{eq:c_pq}
    C^p_q (t) \equiv \langle \Delta T_p (t) \Delta T_q (0) \rangle.
\end{equation}
Solving Eq.~(\ref{eq: LP_nu_unsolved}) for the Laplace-transformed rate matrix gives
\begin{eqnarray}
    \Tilde{\mathcal{R}} (\omega) = \mathcal{C} (0) \cdot \Tilde{\mathcal{C}}^{-1} (\omega) - i \omega \mathcal{I} ,
\end{eqnarray}
\noindent where $\mathcal{I}$ is the identity matrix. Taking the zero frequency limit yields 
\begin{eqnarray}
    \label{eq: gen_GK_w_C0}
    \mathcal{R} = \mathcal{C} (0) \cdot \int^\infty_0 dt\, \mathcal{C}^{-1} (t) .
\end{eqnarray}
Finally, using the definitions of parallel and perpendicular kinetic temperatures, the initial values of the autocorrelations are $C^{ie}_{ie} (0) = C^i_i (0) = C^e_e (0) = 3T^2/N$ and the initial values of the cross-correlations are all zero. Thus, $\mathcal{C}(0) = \left( 3T^2/N \right) \mathcal{I}$, which simplifies Eq.~(\ref{eq: gen_GK_w_C0}) to 
\begin{equation}
    \label{eq: gen_GK}
    \mathcal{R} = \frac{3T^2}{N} \int^\infty_0 dt\, \mathcal{C}^{-1} (t) ,
\end{equation}
where $T$ is the equilibrium temperature and $N$ is the number of particles in the system. Equation~(\ref{eq: gen_GK}) provides the general matrix form of the Green-Kubo relation for temperature relaxation. It captures the influence of cross-correlations and non-Markoffian effects. Using Eq.~(\ref{eq: gen_GK}), all six rates can, in principle, be extracted. In practice, two of the coupling rates, $(\rho^{ie}_i, \rho^i_e)$, could not be resolved because their corresponding correlation functions were below the noise level of MD. Extending the simulation time would reduce statistical noise and enable their extraction, but the MD–GK simulations are already computationally demanding, so magnetization strengths  were limited to
$\beta_e = 8.7$ and lower. In contrast, the less computationally intensive MD–N simulations extended to $\beta_e = 67$.

\subsubsection{Approximate Form\label{sec: simp_GK}}

The general matrix form of the Green-Kubo relation in Eq.~(\ref{eq: gen_GK}) can be simplified in a way that emphasizes the physical effects leading to each relaxation or coupling rate. 
In the weakly magnetized regime, all cross-correlations (off-diagonal terms in the correlation matrix from Eq.~(\ref{eq: corr_matrix})) are zero. Furthermore, for all magnetization strengths considered in this work, the autocorrelations (on-diagonal terms) are much larger than the cross-correlations. 
So, to a good approximation, cross-correlations can be neglected when computing relaxation rates $(\eta)$. As a result, Eq.~(\ref{eq: gen_GK}) reduces to a set of three decoupled scalar Green-Kubo relations
\begin{subequations} \label{eq: simp_ondiag_GK}
\begin{eqnarray}
\eta^{ie} = \frac{3T^2}{N\int^\infty_0 C^{ie}_{ie} (t) dt} \label{eq: simp_ondiag_GK_ie}\\
\eta^{i} = \frac{3T^2}{N\int^\infty_0 C^{i}_{i} (t) dt} \label{eq: simp_ondiag_GK_ii} \\
\eta^{e} = \frac{3T^2}{N\int^\infty_0 C^{e}_{e} (t) dt} . \label{eq: simp_ondiag_GK_ee}
\end{eqnarray}
\end{subequations}
These simplified Green-Kubo expressions were tested by confirming that the coefficients evaluated from them agree with coefficients evaluated from the general Eq.~(\ref{eq: gen_GK}) to within the numerical resolution. 
However, unlike the general form, they reveal that the relaxation rates, $\eta^\alpha$, are primarily determined by the corresponding autocorrelation function $C_\alpha^\alpha(t)$. 
In other words, the cross correlation contributions are small. 

In addition to the relaxation rates, one of the three coupling rates can be obtained because $C^{ie}_e (t)$ and $C^e_{ie} (t)$ are nonzero in the strongly magnetized regime. For all magnetization strengths examined, the remaining cross-correlations were zero within the noise level in the MD simulations $(C^{ie}_i(t) = C^i_{ie}(t) = C^i_e(t) = C^e_i(t) \approx 0)$. Thus, Eq.~(\ref{eq: gen_GK}) reduces to Eq.~(\ref{eq: simp_ondiag_GK_ii}) and a $2 \times 2$ matrix equation. Solving for the off-diagonal rates, we find
\begin{eqnarray} \label{eq: simp_offdiag_GK}
\rho^{ie}_{e} 
&=& \frac{3T^2 S^e_{ie}}{N \left(S^e_{ie} S^{ie}_e - S^{ie}_{ie} S^e_e\right)} \nonumber \\
&=& \frac{3T^2 S^{ie}_e}{N \left(S^e_{ie} S^{ie}_e - S^{ie}_{ie} S^e_e\right)} , 
\end{eqnarray}
where $S^q_p \equiv \int^\infty_0 C^q_p(t)dt$. Our results use the average of these two equations.

\subsection{Simulation Setup}

The MD-GK simulations used nearly the same setup as the MD-N simulations, but with a few key differences. 
After transitioning to an NVE ensemble and turning the magnetic field on, the system continued to be evolved at equilibrium instead of scaling the ion and electron velocities to create a temperature difference. 
In this regard, we note that the Bohr–van Leeuwen theorem holds exactly in equilibrium systems~\cite{Pathria2011Book}, so this avoids the assumption of the small temperature difference having a negligible affect on the evolution that was assumed in the MD-N method.
Beyond this, the main distinction is that the MD-GK simulations were run much longer in time to achieve convergence of the correlation functions. 
Most simulations were run for several hundred thousand electron plasma periods, compared to only a few thousand for the MD-N runs, making the MD-GK simulations substantially more computationally expensive. These longer simulation times were necessary to ensure that the time-correlation functions were well resolved.

The correlation functions from Eq.~(\ref{eq:c_pq}) and in Fig.~\ref{fig: corrs} were computed from the time series of the temperatures, and the rates computed using Eqs.~(\ref{eq: simp_ondiag_GK}) and (\ref{eq: simp_offdiag_GK}). 
However, the decay of $C^{ie}_{ie} (t)$ was much slower than the other autocorrelations, as it is related to the ion-electron relaxation time, which is elongated due to the large mass ratio between ions and electrons. 
To enable a calculation of this, $C^{ie}_{ie} (t)$ was fit to an exponential decay and analytically integrated.

\subsection{Results \label{sec: EMD_rates}}

Results from the MD-GK simulations for $\eta^{ie}$ are shown in Fig.~\ref{fig: nuie00}. 
As with the MD-N simulations, these show that the ion-electron equilibration rate is enhanced at moderate magnetization strengths. 
Specifically, it is flat in the weakly magnetized regime, in agreement with theory, MD-N results, and previous unmagnetized MD studies~\cite{Dimonte_PRL_2008,bernstein2022method}, increases at the transition to strong magnetization, and appears consistent with a plateau at the largest $\beta_e$ value simulated of 8.7. The isotropic reduced model from Eq.~(\ref{eq: unmag_rm}) indicates that $\eta^{ie}$ represents the ion-electron equilibration rate in the regime where $\beta_e \lesssim 5$. Therefore, equilibration is enhanced at these moderate magnetization strengths in agreement with the temperature evolution and relaxation rate results from MD-N shown in Fig.~\ref{fig: MD_theo}. 

Quantifying uncertainty in MD simulations is inherently challenging. For MD-GK, the dominant source of error is expected to be from the finite simulation time. In MD-N, additional uncertainty may result from anomalous partial relaxation during the initial phase. Both methods are also subject to finite-size effects and statistical fluctuations, the latter of which could be reduced by increasing the number of independent runs. The MD-GK rates consistently exceeded the MD-N rates in both the weakly magnetized regime and the strongly magnetized plateau. Given these uncertainties, definitive conclusions about the discrepancies between the MD results and the theoretical predictions are difficult to draw. Nevertheless, both MD methods reproduce the trends predicted by theory.

As shown in Fig.~\ref{fig: nuee_NEMD_TCP}, the model predicts that for magnetization strengths below $\beta_e \lesssim 10$ the total electron isotropization rate $\eta^e$ in an ion–electron plasma (blue curve) exceeds that in a pure electron plasma $\eta^{e,e|e}$ (purple curve). Relaxation rates obtained from MD–GK simulations, together with previous OCP MD data~\cite{Baalrud_PRE_2017,Baalrud_CPP_2017,jose_submitted_2025}, confirm this prediction, indicating that electron isotropization in an ion–electron plasma is driven by both electron–electron and electron–ion collisions. 
Although the Green–Kubo relation in Eq.~(\ref{eq: simp_ondiag_GK_ee}) does not directly yield $\eta^{e,i|e}$, it can be determined by subtracting the electron isotropization rate in a pure electron plasma from the total $\eta^e$ in a two-component plasma.

The electron isotropization rate remains constant in the weakly magnetized regime and decreases exponentially as the system transitions to strong magnetization. This behavior supports the MD–N results, which show that ions cannot effectively exchange energy with the perpendicular degree of freedom of the electrons. The suppression of the electron isotropization rate is so pronounced that it becomes the slowest relaxation process governing the total plasma equilibration, thereby extending the equilibration time by orders of magnitude.

\begin{figure} [!htb] 
\centerline{\includegraphics[width = 8.5cm]{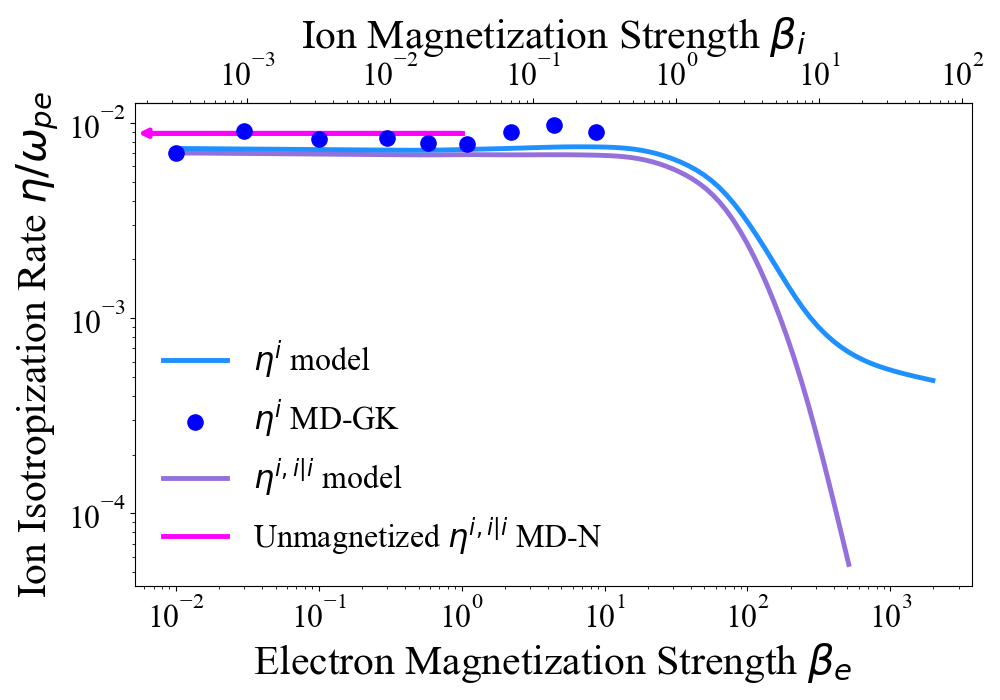}}
\caption { Ion isotropization rate as a function of electron magnetization strength $\beta_e$ (bottom axis) and ion magnetization strength $\beta_i$ (top axis).
The MD–GK results (blue circles) are compared with the theoretical model predictions for ion isotropization $\eta^i$ (blue line). Also shown are the model prediction for ion isotropization due to ion-ion collisions (purple line) and a previous unmagnetized MD study (pink arrow) for ion isotropization in a pure ion plasma~\cite{Baalrud_CPP_2017}.}
  \label{fig: nuii_TCP}
\end{figure}

Figure~\ref{fig: nuii_TCP} shows that ion isotropization rates from MD-GK remain independent of $\beta_e$ as long as the ion magnetization parameter is less than unity ($\beta_i \lesssim 1$), consistent with model predictions. Since the MD-GK simulations were limited to a maximum magnetization strength of $\beta_e=8.7$, or $\beta_i=0.28$, the MD-GK ion isotropization rates remained independent of magnetization strength. Unlike electron isotropization, the total ion isotropization $(\eta^i)$ and the ion isotropization due to ion-ion collisions $(\eta^{i,i|i})$ are nearly identical. The large mass ratio means that electrons transfer little energy to the heavier ions; thus, ion isotropization is primarily driven by ion-ion collisions. 
Although the theory predicts that to change when $\beta_i \gg 1$, the MD simulations were unable to access that regime. 

\begin{figure} 
\centerline{\includegraphics[width = 8.5cm]{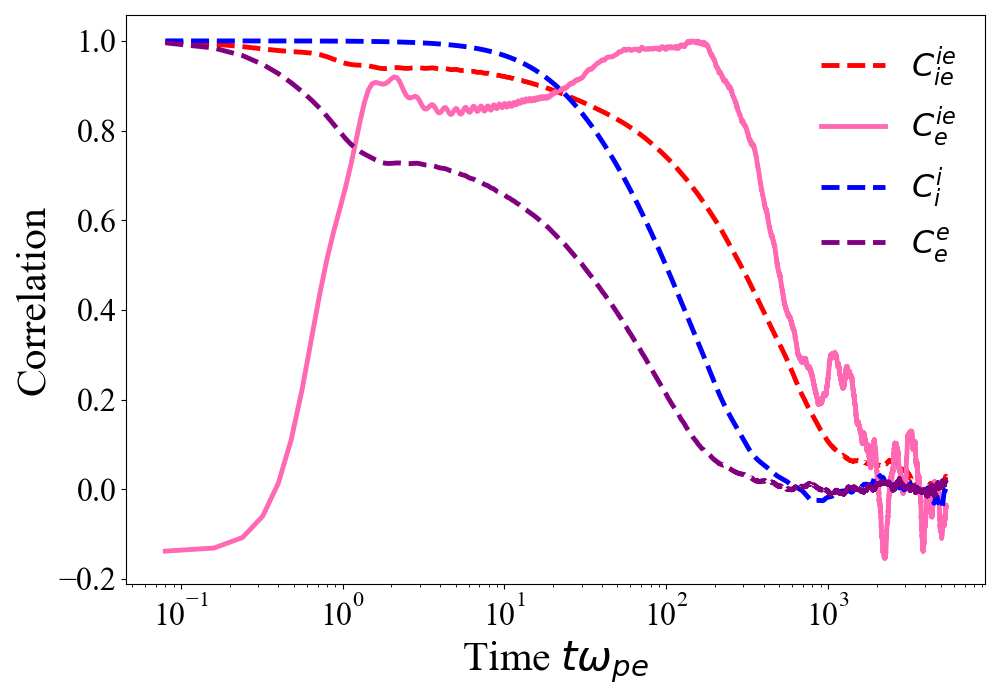}}
\caption {Time correlation functions as a function of time in electron plasma periods at a magnetization strength of $\beta_e=8.7$. Autocorrelations are normalized to their initial value: ion-electron (dashed red), ion (dashed blue), and electron (dashed purple). The cross-correlation between the ion-electron temperature difference and the electron anisotropy is normalized to its maximum value (pink line).}
  \label{fig: corrs}
\end{figure}

Fig.~\ref{fig: corrs} shows examples of the correlation functions used to calculate rates using the Green-Kubo relations in Eqs.~(\ref{eq: gen_GK})-(\ref{eq: simp_offdiag_GK}).
The autocorrelations have a roughly exponential decay as expected. Cross-correlations are different.
Initially, all cross-correlations are essentially zero. Under weak magnetization, they are zero within the noise level of the MD simulations for all time. However, under strong magnetization, some of these cross-correlations, such as $C^{ie}_{e} (t)$, grow in time and eventually decay back to zero.
The existence of these correlations implies that the presence of an ion-electron temperature difference influences the evolution of an electron anisotropy.

\begin{figure} [!htb] 
\centerline{\includegraphics[width = 8.5cm]{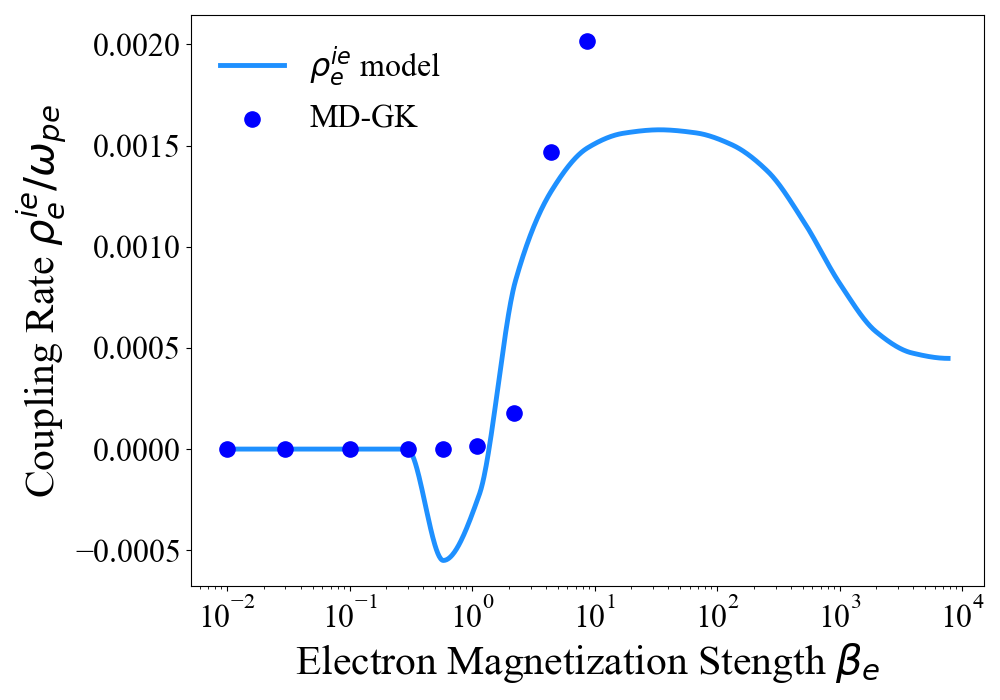}}
\caption {The coupling rate between the ion electron temperature difference and the electron anisotropy $(\rho^{ie}_{e})$ as a function of magnetization strength $(\beta_e)$. The blue curve is the model prediction, and blue circles are MD-GK simulation results computed using Eq.~(\ref{eq: simp_offdiag_GK}).}
  \label{fig: rho_ie_01}
\end{figure}

Results for the coupling rate $\rho^{ie}_{e}$ are shown in Fig.~\ref{fig: rho_ie_01}. This is a quantity that can be computed using the MD-GK, but not using MD-N. The results demonstrate that in the strongly magnetized regime an ion–electron temperature difference generates an electron anisotropy. This coupling rate was found to be positive under strong magnetization. As can be seen in Eq.~(\ref{eq:T_e_evol}), a positive value of $\rho^{ie}_{e}$ indicates that an electron anisotropy $(T_{e\parallel} -T_{e\perp})$ will grow in time if the ions are hotter than the electrons. This behavior can also be observed in the MD-N temperature evolution shown in Figs.~\ref{fig: MD_theo}~(e,f); the initially isotropic temperatures evolve such that the ion and parallel electron temperatures relax more rapidly than the perpendicular electron temperature, thereby creating an electron anisotropy.
It is worth noting that although this process generates an anisotropy, the overall relaxation remains entropy producing and the plasma continues evolving toward a Maxwellian distribution with a common temperature. 

As predicted by the model, $\rho^{ie}_e$ is zero in the weakly magnetized regime, implying that the temperatures relax isotropically. Near the transition to strong magnetization, the MD-GK rate increases, and then plateaus at higher magnetization strength. Although the model predicts a similar trend, it also anticipates a dip to negative values before the rise. This feature is absent in our MD-GK results.
This predicted dip corresponds to an enhanced electron isotropization due to an ion-electron temperature difference, but the effect is expected to have a negligible change in the temperature evolutions in Fig.~\ref{fig: MD_theo} since electron isotropization $(\eta^e)$ is much faster than the predicted negative values of $\rho^{ie}_e$.
Beyond this small disagreement, the MD-GK coupling rates compare well with the model.

\section{Conclusion\label{sec:Conclusion}}

In this work, first-principles MD simulations were utilized to test a recently developed theory of temperature relaxation in strongly magnetized two-component plasmas. The MD-N simulations produced temperature evolution curves that showed good agreement with the theoretical predictions. Relaxation rates extracted from the MD-N and MD-GK data further confirm these predictions. These results demonstrate that at moderate magnetization strengths ($1 \lesssim \beta_e \lesssim 10$) the relaxation time is decreased compared to an unmagnetized system, whereas at high magnetization strengths ($\beta_e \gtrsim 10$), the relaxation time is lengthened by orders of magnitude. The prolonged relaxation time was found to be due to the perpendicular electron temperature relaxation rate exponentially decreasing with magnetic field strength.

This work helps to confirm the new temperature evolution model~\cite{josePRE2025}. This model has several applications, but is particularly relevant to ALPHA’s manipulation of non-neutral, strongly magnetized two-component plasmas. ALPHA regularly works with antiproton-electron and positron-Beryllium plasmas confined in Penning-Malmberg traps, which are required to be strongly magnetized~\cite{Davidson2001Book,Andresen2011,Andresen_PRL_2008,Baker_Nature_Comm_2021, Fajans_POP_2020,Fajans_POP_2020, Wurtele2013}.
An improved understanding of temperature equilibration may help achieve colder plasmas, thus increasing antihydrogen production~\cite{GlinskyONeil1991}.
Future work will investigate how the theoretical rates tested in this work could be experimentally measured in a Penning-Malmberg trap under experimental conditions commonly used at ALPHA.

\section{Data Availability Statement}

The data that support the findings of this study are available from the corresponding author upon reasonable request.

\section{Author Declarations}
\subsection{Conflict of Interest}
The authors have no conflicts to disclose.
 
\begin{acknowledgments}
The authors thank Jarett LeVan and Ryan M. Park for helpful conversations during the preparation of the manuscript.
This material is based upon work supported by NSF grant award Nos.~PHY-2205506, PHY-2205620, and PHY-2512424.   
It used Expanse at San Diego Supercomputer Center through allocation PHY-150018 from the Advanced Cyberinfrastructure Coordination Ecosystem: Services \& Support (ACCESS) program, which is supported by National Science Foundation grants \#2138259, \#2138286, \#2138307, \#2137603, and \#2138296.
\end{acknowledgments}

\appendix

\section{Symmetry of the Correlation Matrix \label{app: C_sym}}

Here, we show that the correlation matrix is symmetric. From that, the rate matrix is shown to also be symmetric and have only 6 independent rates rather than 9. Time-correlation functions are invariant under time translation, which implies
\begin{eqnarray} \label{eq: Hansen_719}
    C^p_q (t) = \epsilon_p \epsilon_q C^p_q (-t) = \epsilon_p \epsilon_q C^q_p (t)
\end{eqnarray}
where $\epsilon_p$ and $\epsilon_q$ are the time-reversal signatures of $\Delta T_p$ and $\Delta T_q$, respectively~\cite{hansen2013theory}. If a temperature difference is time reversible, then its time signature is one. Making use of the kinetic definitions of temperatures,
\begin{subequations}\label{eq: kin_temp_defs}
\begin{eqnarray}
    &&\Delta T_e (t) = \frac{m_e}{2 k_B N} \nonumber \\
    &&\sum^N_{k=1} \left( 2 v^2_{k,ez} (t) - v^2_{k,ex} (t) - v^2_{k,ey} (t)\right) , \\
    &&\Delta T_i (t) = \frac{m_i}{2 k_B N} \nonumber \\
    &&\sum^N_{k=1} \left( 2 v^2_{k,iz} (t) - v^2_{k,ix} (t) - v^2_{k,iy} (t)\right) , \\
    &&\Delta T_{ie} (t) = \frac{1}{3 k_B N}  \Biggl(m_i \sum^N_{j=1} \left( v^2_{j,ix} + v^2_{j,iy} + v^2_{j,iz} \right) \nonumber \\
    &&- m_e \sum^N_{k=1} \left( v^2_{k,ex} + v^2_{k,ey} + v^2_{k,ez} \right) \Biggr)  ,
\end{eqnarray} 
\end{subequations}
where $v_{k,s\alpha}$ is the velocity of the $k$th particle of species $s$ in the Cartesian direction $\alpha$.
Take $t \rightarrow -t$. This implies $\textbf{v}(t) \rightarrow \textbf{v}(-t) = -\textbf{v}(t)$. Using this in the temperature differences results in
\allowdisplaybreaks
\begin{subequations}\label{eq: kin_temp_defs_t_revers}
\begin{eqnarray}
    &&\Delta T_e (t) = \frac{m_e}{2 k_B N} \nonumber \\
    &&\sum^N_{k=1} \left( 2 \left(-v_{k,ez} (t) \right)^2 - \left(-v_{k,ex} (t) \right)^2 - \left(-v_{k,ey} (t) \right)^2 \right)  \\
    &&\Delta T_i (t) = \frac{m_i}{2 k_B N} \nonumber \\
    &&\sum^N_{k=1} \left( 2 \left(-v_{k,iz} (t) \right)^2 - \left(-v_{k,ix} (t) \right)^2 - \left(-v_{k,iy} (t) \right)^2 \right)  \\
    &&\Delta T_{ie} (t) = \frac{1}{3 k_B N}  \\
    &&\Biggl(m_i \sum^N_{j=1} \left( \left(-v_{j,ix} (t) \right)^2 + \left(-v_{j,iy} (t) \right)^2 + \left(-v_{j,iz} (t) \right)^2 \right) \nonumber \\
    &&- m_e \sum^N_{k=1} \left( \left(-v_{k,ex} (t) \right)^2 + \left(-v_{k,ey} (t) \right)^2 + \left(-v_{k,ez} (t) \right)^2  \right) \Biggr)  .\nonumber
\end{eqnarray} 
\end{subequations}
This simplifies to the original definitions in Eq.~(\ref{eq: kin_temp_defs}). Therefore,
\begin{eqnarray}
    \epsilon_e = \epsilon_i = \epsilon_{ie} = 1 .
\end{eqnarray}
From Eq.~(\ref{eq: Hansen_719}), this means $C^p_q (t) = C^q_p (t)$ and therefore the correlation matrix is symmetric. Using the Green-Kubo relation in Eq.~(\ref{eq: gen_GK}), and the fact that the correlation matrix is symmetric, it was found that the rate matrix was also symmetric. Therefore, the rate matrix can be expressed in terms of 6 independent rates rather than 9.

\section{Definition of Rates \label{app: def_rates}}

The definitions of the rates in Eq.~(\ref{eq:T_evolution}) are different from those used in Ref.~\cite{josePRE2025} due to the newfound symmetry which reduces the number of rates from $9$ to $6$. Furthermore, the new definitions provide a more intuitive physical understanding of each rate.
The relation between the new and old rate definitions are
\begin{subequations}\label{eq: rate_defs}
\begin{eqnarray}
    \eta^{ie} &\equiv& 2 \nu^{ie}_{00} 
    = 2 \mu^{ie}_{00} \\
    \eta^{i,i|e} &\equiv& \nu^{i e}_{A10} 
    = - \mu^{i e}_{A10} \\
    \eta^{e,i|e} &\equiv& - n_r \nu^{e i}_{A10} \left(1 + \chi^{ei}_{A10}\right) 
    = - \mu^{ei}_{A10} \\
    \eta^{i,i|i} &\equiv& 3 \nu^{ii} 
    = \mu^{ii} \\
    \eta^{e,e|e} &\equiv& 3 \nu^{ee} 
    = \mu^{ee} \\
    \rho^{ie}_{i} &\equiv& 3 \nu^{ie}_{10}
    = 3 \mu^{ie}_{10}
    = \frac{2}{3} \nu^{ie}_{A00}
    = \frac{2}{3} \mu^{ie}_{A00} \\
    \rho^{ie}_{e} &\equiv& 3 \nu^{i e}_{01} \left(1 + \chi^{ie}_{01}\right)
    = 3 \mu^{ie}_{01} \nonumber \\
    &=&  - \frac{2}{3} n_r \nu^{ e i}_{A00}
    = \frac{2}{3} \mu^{ e i}_{A00} \\
    \rho^{i}_{e} &\equiv& - \nu^{i e}_{A01} \left(1 + \chi^{ie}_{A01}\right) 
    = \mu^{i e}_{A01} \nonumber \\
    &=& n_r \nu^{e i}_{A01} \chi^{ei}_{A01}
    = \mu^{e i}_{A01} .
\end{eqnarray} 
\end{subequations}

\bibliography{references}	

\end{document}